# Indices in XML databases

Hadj Mahboubi

University of Lyon (ERIC Lyon 2)

5 avenue Pierre Mendès-France, 69676 Bron Cedex, France

Phone: +33 478 773 111 — Fax: +33 478 772 375

hadj.mahboubi@eric.univ-lyon2.fr

Jérôme Darmont [*]

University of Lyon (ERIC Lyon 2)

5 avenue Pierre Mendès-France, 69676 Bron Cedex, France

Phone: +33 478 774 403 — Fax: +33 478 772 375

jerome.darmont@univ-lyon2.fr

**Indices in XML databases**

INTRODUCTION

Since XML (eXtensible Markup Language; Bray et al., 2004) emerged as a standard for information representation and exchange, storing, indexing, and querying XML documents have become major issues in database research. Query processing and optimization are very important in this context, and indices are data structures that help enhance performances substantially. Though XML indexing concepts are mainly inherited from relational databases, XML indices bear numerous specificities.

The aim of this article is to present an overview of state-of-the-art XML indices, and to discuss the main issues, tradeoffs and future trends in XML indexing. Furthermore, since XML is gaining importance for representing business data for analytics (Beyer et al., 2005), we also present an index we specifically developed for XML data warehouses.

BACKGROUND

Indexing and querying XML documents through path expressions expressed in XPath (Clark & DeRose, 1999) and XQuery (Boag et al., 2006) have been the focus of many research studies. Two families of approaches aim at efficiently processing path join queries. They are based on structural summaries and numbering schemes, respectively.

Structural summary-based indices

Structural index-based approaches help traverse XML documents' hierarchies by referencing structural information about these documents. These techniques extract structural information directly from data and create a structural summary that is a labeled, directed graph. Graph schemas can be used as indices for path queries. Dataguide (Goldman & Widom, 1997) and 1-index (Milo & Suciu, 1999) belong to this family of indices.

Dataguide's structure describes by one single label all the nodes whose labels (names) are identical. Its definition is based on targeted path sets, i.e., sets of nodes that are reached by traversing a given path.

1-index clusters nodes according to a bisimilarity relationship. Two nodes are said bisimilar if they share identical label paths in the XML data graph. Bisimilar nodes are grouped together into one index node. A 1-index is smaller than the initial data graph and thereby facilitates query evaluation. To help select labels or evaluate path expressions, hash tables or B-trees are used to index graph labels.

Dataguide and 1-index code all paths from the root node. The size of such summary structures may be larger than the original XML document, which degrades query performance. A(k)-index (Kaushik et al., 2002) is a variant of 1-index that is based on k-dissimilarity and builds an approximate index to reduce its graph's size. An A(k)-index can retrieve, without referring to the data graph, path expressions of length of at most k, where k controls index resolution and influences index size in a proportional manner. However, for large values of k, index size may still become very large. For small values of k, index size is substantially smaller, but A(k)-index cannot handle long path expressions.

To accommodate path expressions of various lengths, without unnecessarily increasing index size, D(k)-index (Qun et al., 2003) assigns different values of k to index nodes. These values conform to a given set of frequently-used path expressions (FUPs). Small or large values of k are assigned to index parts that are targeted by short or long path expressions, respectively. To help evaluate path expressions with branching, a variant called UD(k, l)-index (Wu et al., 2003) also imposes downward similarity.

AD(k)-index (He & Yang, 2004) builds a coarser index than A(k)-index, but suffers from over-refinement. M(k)-index, an improvement of D(k)-index, solves the problem of large scan space within the index, without affecting path coverage. However, there is a drawback in this design: M(k)-index requires adapting to a given list of FUPs.

U(*)-index (universal, generic index; Boulos & Karakashian, 2006), like 1-index, exploits bisimilarity. However, U(*)-index exploits a special node labeling scheme to prune the search space and accelerate XPath evaluations. Furthermore, U(*)-index does not need to be adapted to any particular list of FUP; it has a uniform resolution, and is hence more generic.

APEX (Chung et al., 2002) is an adaptive index that searches for a trade-off between size and effectiveness. Instead of indexing all paths from the root, APEX only indexes frequently-used paths and preserves the structure of source data in a tree. However, since FUPs are stored in the index, path query processing is quite efficient. APEX is also workload-aware, i.e., it can be dynamically updated according to changes in query workload. A data mining method is used to extract FUPs from the workload for incremental update (Agrawal & Srikant, 1995).

The main weakness of these indices is that they can only answer single path expressions directly. To process so-called branching path expressions, whose graphical representation contains branches and corresponds to a small tree (or twig), they must perform a costly join operation. To reduce the number of joins, XJoin-index (Bertino et al., 2004) pre-computes some structural semi-join results to support attribute selection, possibly involving several attributes, detection of parent-child relationships, and counting.

Finally, other techniques such as extended inverted lists (Zhang et al., 2001) and Fabric (Cooper et al., 2001) are aimed at processing containment queries over XML data stored in relational databases. Containment queries are based on relationships among elements, attributes and their contents. Extended inverted lists include a text index (T-index; Milo & Suciu, 1999) that is similar to traditional indices in information retrieval systems, and an element index (E-index) that maps elements into inverted lists.

Fabric indexes several XML documents by encoding paths, from root to leaves. The resulting indicators are then inserted into a Patricia trie (Cooper et al., 2001), which processes them like simple strings. A dictionary stores correspondences between indicators and path label names. To use this index, query labels are also transformed into indicators by exploiting the dictionary.

Numbering scheme-based indices

A numbering scheme encodes each XML element by its positional information in its document's hierarchy. Most numbering schemes reported in the literature are based either on a tree-traversal order, or on the textual positions of start and end tags (Srivastava et al., 2002).

If such a numbering scheme is embedded in the labeled trees of XML documents, a structural relationship (e.g., ancestor-descendant) between a pair of elements can be determined quickly without traversing the whole tree.

To evaluate queries involving structural relationships, structural join indices efficiently support functions such as *findDescendants* and *findAncestors* that are needed in structural joins. For instance, a B+-tree may be built on the joining element's *StartPos* attribute (Chien et al., 2002). XR-tree (XML Region Tree; Jiang et al., 2003) is a dynamic external memory index structure that is specifically designed for strictly nested XML data. Actually, an XR-tree is a B+-tree with a complex index key entry and extra stab lists associated with its internal nodes.

XB-tree (Bruno et al., 2002) combines structural features of both B+-tree and R-tree. XB-tree first indexes pre-assigned intervals of elements from a tree structure. Next, it organizes the intervals' starting points as a B+-tree. Each internal node maintains a set of regions that completely includes all regions in their child nodes. Regions in XB-tree nodes may overlap partially.

XML structural join-based experiments performed on these indices indicate that they achieve comparable performances for non-recursive XML data (i.e., XML documents with no node-to-node internal references), while XB-tree outperforms the other indices for highly recursive XML data (Li et al., 2004).

INDICES IN XML DATA WAREHOUSES

XML data warehouses form an interesting basis for decision-support applications that exploit so-called complex data (Darmont et al., 2005). Several studies address the issue of designing and building XML data warehouses. They use XML documents to manage or represent warehouse facts and/or dimensions (Pokorný, 2002; Hümmer et al., 2003; Park et al., 2005). This approach helps store XML documents natively and query them easily with XML query languages. However, decision-support queries are generally complex. They indeed typically involve several join and aggregation operations. In addition, many XML-native DBMSs show relatively poor performances when data volume is very large and queries are complex.

Most existing XML indexing techniques are applicable only onto XML data that are targeted by single path expressions. However, in XML data warehouses, queries are complex and include several path expressions. Furthermore, building existing indices on an XML warehouse causes a loss of information in decision-support query resolution. Indeed, clustering (1-index and variants) or merging (Dataguide) identical labels causes the disappearance of fact-to-dimension relationships, which are essential to process analytical queries. We illustrate this issue in the following example.

Let us consider a sample warehouse document, *cube.xml*, composed of *cell* (fact) elements (Figure 1(a)). Each cell is identified by a combination of dimension identifiers and one or more measures. Figure 1(b) features the corresponding 1-index. Since 1-index represents cells linearly, i.e., all labels sharing the same label path are represented by one label only, the dimension combinations that identify facts are lost.

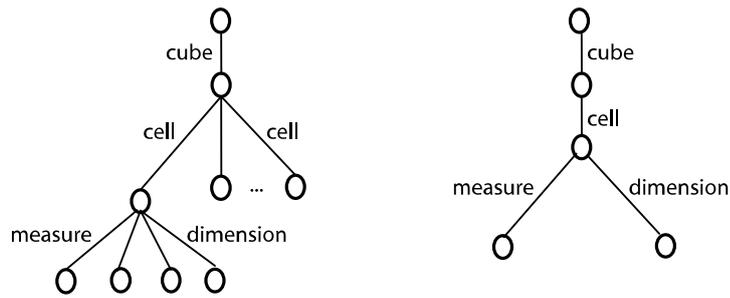

Figure 1: *cube.xml* document structure (a) and corresponding 1-index (b)

Eventually, most of the approaches we presented in the previous section can only index one XML document at a time, whereas in XML warehouses, data are typically stored in several fact and dimension XML documents; and analytic queries must be performed over these documents. Fabric does handle multiple documents, but it is not adapted to XML data warehouses either, because it does not take into account relationships between XML documents (i.e., fact-to-dimension references in our case).

To address the issues of multiple path expressions in analytic queries, loss of referential information and multi-document indexing, we have proposed a new index that is specifically adapted to XML, multidimensional data warehouses (Mahboubi et al., 2006a). This data structure help optimize access time to several XML documents by eliminating join costs, while preserving information contained in the initial warehouse.

To implement our indexing strategy, we selected XCube (Hümmer et al., 2003) as a reference data warehouse model. Since other XML warehouse models from the literature are relatively similar, this is not a binding choice. XCube's advantage is its simple structure for representing facts and dimensions in a star schema: one document for facts (*facts.xml*) and another one for all dimensions (*dimensions.xml*).

Our index structure is designed to preserve relationships between facts and dimensions. To achieve this goal, we move data from *facts.xml* and *dimensions.xml* into a common structure that actually is our index. This process helps store facts, dimensions and their attributes into the same XML element. It wholly eliminates join operations, since all necessary information for a join operation is stored in the same index cell. This data structure is also stored into an XML document (*index.xml*). Queries need to be rewritten to exploit our index instead of the initial warehouse. This rewriting process consists in preserving selection and aggregation operations, while eliminating joins.

To validate our proposal, we performed both a complexity study and field experiments. Our tests showed that using our index structure significantly improved the response time of a typical decision-support query expressed in XQuery. Furthermore, they also demonstrated that well-indexed XML-native DBMSs could compete with relational, XML-compatible DBMSs.

FUTURE TRENDS

As we underlined in the background section, structural summary approaches generate large indices and do not support partial path matching queries. Labeling schemes allow to quickly identify relationships among element nodes and to reduce index size, but fail to support dynamic XML data. Furthermore, the semi-structured nature of XML data and requirements on query flexibility pose unique challenges to indexing methods. Hence, quite recently, researchers proposed hybrid indexing techniques (Catania et al., 2006). XML indexing is likely to keep on following this path, while more specific solutions may also appear, e.g., for XML data warehouses.

The XML warehouse index structure we propose also suffers from these common weaknesses (index size and construction cost). We indeed merge all warehouse data into the same structure. In addition, this process needs to parse all elements within the warehouse XML documents. Index construction is thus very costly. FUPs proposed by Min et al. (2005) might be a solution. FUPs are obtained from a representative workload with data mining approaches, and represent frequent join operations. This could help us materialize these operations only within our index structure.

More generally, XML indexing strategies should be better-integrated in host XML-native DBMSs. This would certainly help develop incremental strategies for index maintenance. Moreover, in our particular case, query rewriting mechanisms would also be more efficient if they were part of the system.

Finally, it is crucial to carry on adapting or developing highly efficient optimization techniques in XML-native DBMSs and relational, XML-enabled DBMSs. Several interesting leads are currently being researched, such as XML view materialization (Mahboubi et al., 2006b; Phillips et al., 2006) or partitioning (Bonifati et al., 2006).

CONCLUSION

Neither XML-native nor XML-enabled DBMSs implement most of the indexing techniques presented in this article. Both classes of systems indeed support only basic solutions.

Relational, XML-enabled DBMSs use simple structural indices such as B-trees and their derivatives. Similarly, most XML-native DBMSs index only element and attribute contents and tag names. In both cases, either full-text inverted indices for indexing textual contents, or path indices are typically adopted. Hence, in conclusion, we strongly believe that XML DBMSs should now feature state-of-the-art, XML-specific indexing schemes to be able to compete with relational DBMSs in terms of performance.

TERMS AND DEFINITIONS

Database management system (DBMS): Software set that handles structuring, storage, maintenance, update and querying of data stored in a database.

XML-enabled DBMS: Database system in which XML data may be stored and queried from relational tables. Such a DBMS must either map XML data into relations and translate queries into SQL, or implement a middleware layer allowing native XML storing and querying.

XML-native DBMS (NXD): Database system in which XML data are natively stored and queried as XML documents. An NXD provides XML schema storage and implements an XML query engine (typically supporting XPath and XQuery). eXist (Meier, 2002) and X-Hive (X-Hive Corporation, 2007) are examples of NXDs.

XML data warehouse: XML database that is specifically modeled (i.e., multidimensionally, with a star-like schema) to support XML decision-support and analytic queries.

Index: Physical data structure that allows direct (vs. sequential) access to data and thereby considerably improves data access time.

Structural summary-based index: Labeled-graph structure that summarizes XML graph structural information.

Numbering scheme-based index: Tree structure in which each XML data node is uniquely identified by an interval.